\documentclass[12pt]{article}

\def\square{\kern1pt\vbox{\hrule height 1.2pt\hbox{\vrule width 1.2pt\hskip 3pt
   \vbox{\vskip 6pt}\hskip 3pt\vrule width 0.6pt}\hrule height 0.6pt}\kern1pt}

\begin{document}

\begin{titlepage}

\begin{flushright}
CECS-PHY-10/01 \\ CCTP-2010-1 \\ UFIFT-QG-10-01
\end{flushright}

\vskip 1cm

\begin{center}
{\bf De Sitter Breaking through Infrared Divergences}
\end{center}

\vskip .5cm

\begin{center}
S. P. Miao$^*$
\end{center}

\begin{center}
\it{Centro de Estudios Cientificos (CECS) \\
Casilla 1469, Valdivia, CHILE}
\end{center}

\vskip .3cm

\begin{center}
N. C. Tsamis$^{\dagger}$
\end{center}

\begin{center}
\it{Department of Physics, University of Crete \\
GR-710 03 Heraklion, HELLAS}
\end{center}

\vskip .3cm

\begin{center}
R. P. Woodard$^{\ddagger}$
\end{center}

\begin{center}
\it{Department of Physics, University of Florida \\
Gainesville, FL 32611, UNITED STATES}
\end{center}

\vspace{.3cm}

\begin{center}
ABSTRACT
\end{center}
Just because the propagator of some field obeys a de Sitter invariant
equation does not mean it possesses a de Sitter invariant solution.
The classic example is the propagator of a massless, minimally
coupled scalar. We show that the same thing happens for massive scalars
with $M_S^2 < 0$, and for massive transverse vectors with $M_V^2 \leq 
-2 (D\!-\!1) H^2$, where $D$ is the dimension of spacetime and $H$ is 
the Hubble parameter. Although all masses in these ranges give infrared 
divergent mode sums, using dimensional regularization (or any other 
analytic continuation technique) to define the mode sums leads to the 
incorrect conclusion that de Sitter invariant solutions exist except at
discrete values of the masses.

\begin{flushleft}
\end{flushleft}

\begin{flushleft}
$^*$ e-mail: smiao@cecs.cl \\
$^{\dagger}$ e-mail: tsamis@physics.uoc.gr \\
$^{\ddagger}$ e-mail: woodard@phys.ufl.edu
\end{flushleft}

\end{titlepage}

\section{Introduction}

Dimesnional regularization \cite{dimreg} is a wonderful tool for 
perturbative computations in quantum field theory but, like all
tools, it can be misused. One way for this to happen involves the
infamous {\it automatic subtraction}: dimensional regularization 
registers only logarithmic divergences; it sets power law 
divergences to zero. When dimensional regularization is employed 
to control an ultraviolet divergence the automatic subtraction
is no problem because the right way to deal with ultraviolet
divergences is by subtracting them with counterterms. In that 
case the automatic subtraction merely spares one the labor of 
explicitly working out the counterterms required to subtract off
any power law divergences. However, dimensional regularization 
can also be used to control infrared divergences, and this 
leads to errors if one fails to recall that the technique
automatically sets power law divergences to zero.

The appearance of an infrared divergence in the answer to a
quantum field theoretic question means that something about
the question is unphysical. The right thing to do in that case
is to revise the question so as to make it more physical. The 
classic example of this is the infrared divergences one encounters 
when computing exclusive scattering amplitudes in quantum 
electrodynamics. Because the photon is massless and all real 
detectors have finite resolution, one can never exclude the 
possibility that the final state contains an extra, very low
energy photon. Including arbitrary numbers of soft photons in the
final state eliminates the infrared divergence \cite{BN}. Another
example is when the vacuum decays, as it does for a massless
scalar with a cubic interaction. Veneziano showed that even 
inclusive scattering amplitudes harbor infrared divergences for 
this system \cite{GV}. The problem in this case is that the free 
scalar vacuum centered around $\phi = 0$ decays, so one cannot 
assume the final state is even stationary, much less centered 
about $\phi = 0$. Infrared finite results can be obtained by 
instead releasing the system at a finite time in a prepared state 
centered around $\phi = 0$ and then following its evolution 
\cite{TW1}.

A peculiar situation arises in curved backgrounds when parameters 
of the background geometry change the infrared properties of the 
particle. For example, consider a homogeneous, isotropic and 
spatially flat background,
\begin{equation}
ds^2 = -dt^2 + a^2(t) d\vec{x} \!\cdot\! d\vec{x} \; .
\end{equation}
The Hubble parameter $H(t)$ and the deceleration parameter $q(t)$ are,
\begin{equation}
H \equiv \frac{\dot{a}}{a} \qquad , \qquad q \equiv -1 - 
\frac{\dot{H}}{H^2} \; .
\end{equation}
The spatial plane wave mode functions for a massless, minimally 
coupled scalar are quite complicated for general $q(t)$ \cite{TW2}
but they take a simple form when $q(t)$ is any constant $q_1$,
\begin{equation}
u(t,k) = \sqrt{\frac{-\pi}{4 q_1 H a}} \, a^{1-\frac{D}2} 
H^{(1)}_{\nu}\Bigl(\frac{-k}{q_1 H a}\Bigr) \qquad {\rm where} \qquad 
\nu = \frac12 - \Bigl(\frac{D \!-\! 2}{2 q_1}\Bigr) \; .
\end{equation}
The naive mode sum for the scalar propagator between $(t,\vec{x})$ and 
$(t',\vec{x}')$ is,
\begin{equation}
\int \!\! \frac{d^{D-1}k}{(2\pi)^{D-1}} \, e^{i \vec{k} \cdot (\vec{x}
- \vec{x}')} \Biggl\{ \theta(t \!-\! t') u(t,k) u^*(t',k)
+ \theta(t' \!-\! t) u^*(t,k) u(t',k) \Biggr\} \; . \label{naive}
\end{equation}
From the small $k$ behavior of the mode functions,
\begin{eqnarray}
-1 \leq q_1 < 0 & \!\!\! \Longrightarrow \!\!\! & 
u u^* =  \frac{4^{\nu-1} \Gamma^2(\nu)}{\pi (-q_1)^{1-2\nu}}
\frac{(a a')^{\nu -\frac{D-1}2}}{(H H^{'})^{\frac12 -\nu}} \times
\frac1{k^{2\nu}} \Biggl\{ 1 \!+\! O(k^2)\Biggr\} , \label{accel} \\
0 < q_1 & \!\!\! \Longrightarrow \!\!\! & 
u u^* =  \frac{4^{-\nu-1} \Gamma^2(-\nu)}{\pi (-q_1)^{1+2\nu}}
\frac{(a a')^{-\nu -\frac{D-1}2}}{(H H^{'})^{\frac12 +\nu}} \times
\frac1{k^{-2\nu}} \Biggl\{ 1 \!+\! O(k^2)\Biggr\} , \qquad \label{decel}
\end{eqnarray}
one can easily recognize that the naive mode sum (\ref{naive}) possesses 
infrared divergences for all values of $q_1$ in the range \cite{FP,JMPW1},
\begin{equation}
-1 \leq q_1 \leq \Bigl(\frac{D \!-\! 2}{D}\Bigr) \qquad \Longrightarrow
\qquad {\rm Infrared\ Divergences} \; . \label{bad}
\end{equation}
However, most of these infrared divergences are of the power law type
which are set to zero by dimensional regularization or any other analytic
continuation method. One only encounters {\it logarithmic} infrared
divergences (either in the leading term or one of the $k^{\mp 2\nu + 2N}$ 
corrections) for the cases \cite{JMPW1},
\begin{eqnarray}
-1 \leq q_1 < 0 & \Longrightarrow & q_1 = - \Bigl( \frac{D \!-\! 2}{D \!-\! 
2 \!+\! 2 N} \Bigr) \qquad {\rm for\ Log\ Divergence} \; , \label{bad1} \\
0 < q & \Longrightarrow & q_1 = \Bigl( \frac{D \!-\!2}{D \!+\! 2N} \Bigr) 
\qquad {\rm for\ Log\ Divergence} \; . \qquad \label{bad2}
\end{eqnarray}

If one were to incorrectly employ dimensional regularization (or any 
other analytic continuation method) to control the infrared divergence
of the naive mode sum (\ref{naive}) it would seem to give a finite
result for the propagator except at the discrete $q_1$ values 
(\ref{bad1}-\ref{bad2}) for which there happens to be a logarithmic
infrared divergence \cite{early,recent}. In fact the mode sum is
infrared divergent for all values of $q_1$ in the range (\ref{bad}).
The right way of dealing with these infrared divergences is not
to subtract them but rather to correct whatever unphysical assumption 
about the system produced them. In this case the problem derives from 
incorrectly assuming that all modes of the initial state are in 
coherent Bunch-Davies vacuum, even the ones with super-horizon 
wavelengths which cannot be controlled by a local observer. The
system can be made infrared finite either by starting with the
super-horizon modes in some less singular vacuum \cite{AV}, or else 
by working on a spatially compact manifold which has no initially 
super-horizon modes \cite{TW3}. Both of these procedures augment the 
naive propagator (\ref{naive}) with extra terms which can mediate 
important effects \cite{JMPW2}.

The purpose of this paper is to point out that a similar situation exists
on de Sitter background ($a(t) = e^{H t}$, with $H$ constant) if one 
considers different values of the mass-squared $M^2$. This has important 
consequences for the construction of de Sitter invariant propagators. We 
show that the formally de Sitter invariant mode sums are infrared singular 
for minimally coupled 
scalars with $M_S^2 \leq 0$, and for transverse vectors with $M_V^2 \leq 
-2(D\!-\!1) H^2$. However, one only encounters logarithmic infrared 
divergences for the special cases,
\begin{equation}
M_S^2 = -N (D \!-\! 1 \!+\! N) H^2 \qquad {\rm and} \qquad 
M_V^2 = -(N \!+\! 2) (D \!-\! 1 \!+\! N) H^2 \; . \label{probs}
\end{equation}
Using dimensional regularization (or any other analytic continuation 
technique) to evaluate the naive mode sums leads to the incorrect 
conclusion that a de Sitter invariant propagator exists, except for
the ``problematic'' cases (\ref{probs}). The correct result is rather
that the naive mode sums diverge for any scalar with $M_S^2 \leq 0$
and for any transverse vector with $M_V^2 \leq -2(D \!-\!1) H^2$. We
make the system infrared finite by working on the compact spatial
manifold $T^{D-1}$, and we obtain explicit results for the leading
infrared corrections to the propagator. These corrections break de 
Sitter invariance, just as has long been known occur for the massless, 
minimally coupled scalar \cite{AF}.

In section 2 we review our conventions for the de Sitter geometry.
Section 3 treats minimally coupled scalars, and section 4 is devoted
to transverse vectors. Section 5 summarizes and discusses our results.

\section{The de Sitter Geometry}

We work on the open conformal coordinate submanifold of $D$-dimensional de
Sitter space. A spacetime point $x^{\mu}$ can be decomposed into its temporal
($x^0$) and spatial $x^i$ components which take values in the ranges,
\begin{equation}
-\infty < x^0 < 0 \qquad {\rm and} \qquad {\rm and} -\infty < x^i <
+\infty \; .
\end{equation}
In these coordinates the invariant element is,
\begin{equation}
ds^2 \equiv g_{\mu\nu} dx^{\mu} dx^{\nu} = a_x^2 \eta_{\mu\nu} dx^{\mu}
dx^{\nu} \; ,
\end{equation}
where $\eta_{\mu\nu}$ is the Lorentz metric and $a_x = -1/Hx^0$ is the
scale factor. The parameter $H$ is known as the ``Hubble constant''.

Most of the various propagators between points $x^{\mu}$ and
$z^{\mu}$ can be expressed in terms of the de Sitter length function
$y(x;z)$,
\begin{equation}
y(x;z) \equiv a_x a_z H^2 \Biggl[ \Bigl\Vert \vec{x} \!-\! \vec{z}
\Bigr\Vert^2 - \Bigl(\vert x^0 \!-\! z^0\vert \!-\! i \epsilon\Bigr)^2 
\Biggr]\; . \label{ydef}
\end{equation}
Except for the factor of $i\epsilon$ (whose purpose is to enforce Feynman
boundary conditions) the function $y(x;z)$ is closely related to the
invariant length $\ell(x;z)$ from $x^{\mu}$ to $z^{\mu}$,
\begin{equation}
y(x;z) = 4 \sin^2\Bigl( \frac12 H \ell(x;z)\Bigr) \; .
\end{equation}

Because $y(x;z)$ is a de Sitter invariant, so too are covariant 
derivatives of it. With the metrics $g_{\mu\nu}(x)$ and $g_{\mu\nu}(z)$, 
the first three derivatives of $y(x;z)$ furnish a convenient basis of
de Sitter invariant bi-tensors \cite{KW1},
\begin{eqnarray}
\frac{\partial y(x;z)}{\partial x^{\mu}} & = & H a_x \Bigl(y \delta^0_{\mu}
\!+\! 2 a_z H \Delta x_{\mu} \Bigr) \; , \\
\frac{\partial y(x;z)}{\partial z^{\nu}} & = & H a_z \Bigl(y \delta^0_{\nu}
\!-\! 2 a_x H \Delta x_{\nu} \Bigr) \; , \\
\frac{\partial^2 y(x;z)}{\partial x^{\mu} \partial z^{\nu}} & = & H^2 a_x a_z
\Bigl(y \delta^0_{\mu} \delta^0_{\nu} \!+\! 2 a_z H \Delta x_{\mu}
\delta^0_{\nu} \!-\! 2 a_x \delta^0_{\mu} H \Delta x_{\nu} \!-\! 2
\eta_{\mu\nu}\Bigr) \; . \qquad
\end{eqnarray}
Here and subsequently $\Delta x_{\mu} \equiv \eta_{\mu\nu} (x \!-\! z)^{\nu}$.
Acting more covariant derivatives just gives back the basis tensors, for
example \cite{KW1},
\begin{equation}
\frac{D^2 y(x;z)}{Dx^{\mu} Dx^{\nu}} = H^2 (2 \!-\!y) g_{\mu\nu}(x) \qquad ,
\qquad \frac{D^2 y(x;z)}{Dz^{\mu} Dz^{\nu}} = H^2 (2 \!-\!y) g_{\mu\nu}(z) \; .
\end{equation}
Similarly, the contraction of any pair of the basis tensors produces 
more basis tensors \cite{KW1},
\begin{eqnarray}
g^{\mu\nu}(x) \frac{\partial y}{\partial x^{\mu}} \frac{\partial y}{\partial
x^{\nu}} & = & H^2 \Bigl(4 y - y^2\Bigr) = g^{\mu\nu}(z) \frac{\partial y}{
\partial z^{\mu}} \frac{\partial y}{\partial z^{\nu}} \; , \\
g^{\mu\nu}(x) \frac{\partial y}{\partial x^{\nu}} \frac{\partial^2 y}{
\partial x^{\mu} \partial z^{\sigma}} & = & H^2 (2-y) \frac{\partial y}{
\partial z^{\sigma}} \; , \\
g^{\rho\sigma}(z) \frac{\partial y}{\partial z^{\sigma}}
\frac{\partial^2 y}{\partial x^{\mu} \partial z^{\rho}} & = & H^2 (2-y)
\frac{\partial y}{\partial x^{\mu}} \; , \\
g^{\mu\nu}(x) \frac{\partial^2 y}{\partial x^{\mu} \partial z^{\rho}}
\frac{\partial^2 y}{\partial x^{\nu} \partial z^{\sigma}} & = & 4 H^4
g_{\rho\sigma}(z) - H^2 \frac{\partial y}{\partial z^{\rho}}
\frac{\partial y}{\partial z^{\sigma}} \; , \\
g^{\rho\sigma}(z) \frac{\partial^2 y}{\partial x^{\mu}\partial z^{\rho}}
\frac{\partial^2 y}{\partial x^{\nu} \partial z^{\sigma}} & = & 4 H^4
g_{\mu\nu}(x) - H^2 \frac{\partial y}{\partial x^{\mu}} \frac{\partial y}{
\partial x^{\nu}} \; .
\end{eqnarray}

\section{Scalars}

The purpose of this section is to demonstrate that the de Sitter
invariant propagator equation has no de Sitter invariant solution for 
a scalar with $M_s^2 \leq 0$, and then to construct the leading de 
Sitter breaking correction terms. We begin by giving the propagator 
equation and the plane wave mode functions for Bunch-Davies vacuum. 
The associated mode sum can be evaluated formally to give a de Sitter
invariant hypergeometric function which diverges for isolated values 
of the scalar mass. By studying the infrared behavior of the mode sum 
we show that these isolated values are those for which one of the 
infrared divergences, which are actually present for all $M_s^2 \leq 0$, 
happens to become logarithmic. Except at the isolated values of $M_S^2$, 
all the infrared divergences are of the power law type which dimensional 
regularization (or any analytic continuation method) incorrectly sets 
to zero. We fix the problem by working on a compact spatial manifold 
which has no initially super-horizon modes, and we derive the leading 
infrared corrections.

The propagator of a minimally coupled scalar with mass $M_S$ 
obeys the equation,
\begin{equation}
\sqrt{-g(x)} \, \Bigl[ \square_x - M_S^2\Bigr] i\Delta(x;z) 
= i \delta^D(x \!-\! z) \; . \label{nuprop}
\end{equation}
The plane wave mode function corresponding to Bunch-Davies vacuum is,
\begin{equation}
u(x^0,k) \equiv \sqrt{\frac{\pi}{4 H}} \; a_x^{-\frac{D-1}2} \, 
H^{(1)}_{\nu}(-k x^0) \quad {\rm where} \quad \nu = 
\sqrt{\Bigl(\frac{D \!-\!1}2\Bigr)^2 \!-\! \frac{M_S^2}{H^2}} \; . \label{unu}
\end{equation}
The Fourier mode sum for the propagator on infinite space is \cite{JMPW1},
\begin{eqnarray}
\lefteqn{i\Delta^{\rm dS}(x;z) = \int \!\! \frac{d^{D-1}k}{(2\pi)^{D-1}} \, 
e^{i \vec{k} \cdot (\vec{x} - \vec{z})} \Biggl\{ \theta(x^0 \!-\! z^0) 
u(x^0,k) u^*(z^0,k) } \nonumber \\
& & \hspace{6.4cm} + \theta(z^0 \!-\! x^0) u(x^0,k) u(z^0,k) \Biggr\} . 
\qquad \label{modesum}
\end{eqnarray}
When this sum exists the result is de Sitter invariant \cite{CT},
\begin{eqnarray}
\lefteqn{i\Delta^{\rm dS}(x;z) } \nonumber \\
& & \hspace{-.5cm} = \frac{H^{D-2}}{(4\pi)^{\frac{D}2}} 
\frac{\Gamma(\frac{D-1}2 \!+\! \nu) \Gamma(\frac{D-1}2 \!-\! \nu)}{
\Gamma(\frac{D}2)} \, \mbox{}_2 F_1\Bigl( \frac{D-1}2 \!+\! \nu,
\frac{D-1}2 \!-\! \nu; \frac{D}2;1 \!-\! \frac{y}4\Bigr) \; , \qquad \\
& & \hspace{-.5cm} = \frac{H^{D-2} \Gamma(\frac{D}2 \!-\! 1)}{(4 
\pi)^{\frac{D}2}} \Biggl\{ \Bigl( \frac{4}{y}\Bigr)^{\frac{D}2 -1} 
\mbox{}_2 F_1\Bigl(\frac12 \!+\!\nu,\frac12 \!-\! \nu ; 2 \!-\! 
\frac{D}2 ; \frac{y}4\Bigr) \qquad \nonumber \\
& & + \frac{\Gamma(\frac{D-1}2 \!+\!\nu) \Gamma(\frac{D-1}2 \!-\!
\nu) \Gamma(1 \!-\! \frac{D}2)}{\Gamma(\frac12 \!+\! \nu)
\Gamma(\frac12 \!-\! \nu) \Gamma(\frac{D}2 \!-\! 1)} \, 
\mbox{}_2 F_1\Bigl(\frac{D-1}2 \!+\! \nu , \frac{D-1}2 \!-\! \nu ;
\frac{D}2 ; \frac{y}4\Bigr) \Biggr\} , \qquad \\
& & \hspace{-.5cm} = \frac{H^{D-2}}{(4 \pi)^{\frac{D}2}} \Biggl\{ 
\Gamma\Bigl(\frac{D}2 \!-\! 1\Bigr) \Bigl(\frac{4}{y}\Bigr)^{\frac{D}2-1} 
\nonumber \\
& & \hspace{2cm} - \frac{\Gamma(\frac{D}2) \Gamma(1 \!-\! \frac{D}2)}{
\Gamma(\frac12 \!+\! \nu) \Gamma(\frac12 \!-\! \nu)} \sum_{n=0}^{\infty} 
\Biggl[ 
\frac{\Gamma(\frac32 \!+\! \nu \!+\! n) \Gamma(\frac32 \!-\! \nu \!+\! n)}{
\Gamma(3 \!-\! \frac{D}2 \!+\! n) \, (n \!+\! 1)!} \Bigl(\frac{y}4
\Bigr)^{n - \frac{D}2 +2} \nonumber \\
& & \hspace{5cm} - \frac{\Gamma(\frac{D-1}2 \!+\! \nu \!+\! n)
\Gamma(\frac{D-1}2 \!-\! \nu \!+\! n)}{\Gamma(\frac{D}2 \!+\! n) \,
n!} \Bigl(\frac{y}4\Bigr)^n \Biggr] \Biggr\} . \qquad \label{expansion}
\end{eqnarray}

The gamma function $\Gamma(\frac{D-1}2 -\nu + n)$ on the final line of
(\ref{expansion}) diverges for,
\begin{equation}
\nu = \Bigl(\frac{D\!-\!1}{2}\Bigr) + N \qquad \Longleftrightarrow \qquad
M_S^2 = - N(D \!-\!1 \!+\! N) H^2 \; . \label{logprob}
\end{equation}
Its origin can be understood by performing the angular integration in the
naive mode sum (\ref{modesum}) and then changing to the dimensionless
variable $\tau \equiv k/H\sqrt{a_x a_z}$,
\begin{eqnarray}
\lefteqn{ i\Delta^{\rm dS}(x;z) = \frac{ (a_x a_z)^{-(\frac{D-1}2)}}{
2^D \pi^{\frac{D-3}2} H} \int_0^{\infty} \!\! dk \, k^{D-2}
\Bigl( \frac12 k \Delta x\Bigr)^{-(\frac{D-3}2)} 
J_{\frac{D-3}2}(k \Delta x) } \nonumber \\
& & \times \Biggl\{ \theta(x^0 \!-\! z^0) H_{\nu}^{(1)}(-k x^0) 
H_{\nu}^{(1)}(-k z^0)^* + \theta(z^0 \!-\! x^0) \Bigl({\rm conjugate}
\Bigr) \Biggr\} , \qquad \\
& & \hspace{-.5cm} = \frac{H^{D-2}}{2^D \pi^{\frac{D-3}2}} \int_0^{\infty} 
\!\! d\tau \, \tau^{D-2} \Bigl( \frac12 \sqrt{a_x a_z} \, H \Delta x \tau
\Bigr)^{-(\frac{D-3}2)} J_{\frac{D-3}2}\Bigl( \sqrt{a_x a_z} \, 
H \Delta x \tau\Bigr) \nonumber \\
& & \times \Biggl\{ \theta(x^0 \!-\! z^0) H_{\nu}^{(1)}\Bigl(
\sqrt{\frac{a_z}{a_x}} \, \tau\Bigr) H_{\nu}^{(1)}\Bigl(\sqrt{\frac{a_x}{a_z}} 
\, \tau\Bigr)^* + \theta(z^0 \!-\! x^0) \Bigl({\rm conjugate}\Bigr) 
\Biggr\} . \qquad \label{integral}
\end{eqnarray}
In these and subsequent expressions we define $\Delta x \equiv \Vert
\vec{x} \!-\! \vec{z}\Vert$. That the divergence at (\ref{logprob}) 
is infrared can be seen from the small argument expansion of the Bessel 
function and from its relation to the Hankel function,
\begin{eqnarray}
J_{\nu}(x) & = & \sum_{n=0}^{\infty} \frac{ (-1)^n (\frac12 x)^{\nu +2n}}{
n! \Gamma(\nu \!+\! n \!+\! 1)} \; , \\
H_{\nu}^{(1)}(x) & = & \frac{i \Gamma(\nu) \Gamma(1 \!-\! \nu)}{\pi} 
\Bigl\{ e^{-i\nu \pi} J_{\nu}(x) \!-\! J_{-\nu}(x)\Bigr\} \; .
\end{eqnarray}
The small $\tau$ behavior of the integrand (\ref{integral}) derives
from three factors, the first of which is $\tau^{D-2}$. The second
factor from the Bessel function takes the form,
\begin{equation}
\Bigl( \frac12 \sqrt{a_x a_z} \, H \Delta x \tau\Bigr)^{-(\frac{D-3}2)} 
J_{\frac{D-3}2}\Bigl( \sqrt{a_x a_z} \, H \Delta x \tau\Bigr) =
\frac1{\Gamma(\frac{D-1}2)} \sum_{n=0}^{\infty} C_1(n) \tau^{2n} \; .
\end{equation}
And the final factor from the Hankel functions is,
\begin{equation}
H_{\nu}^{(1)}\Bigl(\sqrt{\frac{a_z}{a_x}} \, \tau\Bigr) 
H_{\nu}^{(1)}\Bigl(\sqrt{\frac{a_x}{a_z}} \, \tau\Bigr)^* =
\frac{2 \Gamma(\nu) \Gamma(2 \nu)}{\pi^{\frac32} \Gamma(\nu \!+\! \frac12) 
\tau^{2\nu} } \sum_{n=0}^{\infty} C_2(n) \tau^{2n} \; .
\end{equation}
One does not need the coefficients $C_1(n)$ and $C_2(n)$ to see that the
small $\tau$ expansion of the integrand takes the form,
\begin{eqnarray}
\lefteqn{\tau^{D-2} \times \frac1{\Gamma(\frac{D-1}2)} \sum_{k=0}^{\infty} 
C_1(k) \tau^{2k} \times \frac{\Gamma^2(\nu) 2^{2\nu}}{\pi^2 \tau^{2\nu} } 
\sum_{\ell=0}^{\infty} C_2(\ell) \tau^{2\ell} } \nonumber \\
& & \hspace{4cm} = \frac{2 \Gamma(\nu) \Gamma(2\nu)}{\pi^{\frac32}
\Gamma(\frac{D-1}2) \Gamma(\nu \!+\! \frac12)} \, \tau^{D-2-2\nu} 
\sum_{n=0}^{\infty} C_3(n) \tau^{2n} \; . \qquad \label{IR}
\end{eqnarray}
Hence the naive mode sum (\ref{modesum}) is infrared divergent for
\begin{equation}
D - 2 - 2\nu \leq -1 \qquad \Longleftrightarrow \qquad M_S^2 \leq 0 \; .
\label{truediv}
\end{equation}
However, there will only be a {\it logarithmic} infrared divergence,
either from the leading term in (\ref{IR}) or from one of the series
corrections at $n = N$, if one has,
\begin{equation}
D - 2 - 2\nu + 2N = -1 \qquad \Longleftrightarrow \qquad M_S^2 = -N
(D \!-\! 1 \!+\! N) H^2 \; .
\end{equation}
This is precisely the condition (\ref{logprob}) for the formal, de Sitter 
invariant mode sum (\ref{expansion}) to diverge.

As emphasized in the Introduction, the appeance of an infrared 
divergence signals that something is unphysical about the quantity 
being computed. The correct response to an infrared divergence is
not to subtract it off, either explicitly or implicitly with the
automatic subtraction of some analytic regularization technique.
One must instead understand the physical problem which caused the
divergence and then to fix the problem.

The divergence (\ref{truediv}) occurs because of the way the 
Bunch-Davies mode functions (\ref{unu}) depend upon $k$ for small $k$. 
The unphysical thing about having Bunch-Davies vacuum for arbitrarily 
small $k$ is that no experimentalist can causally enforce it (or any
other condition) for super-horizon modes. This has led to two fixes:
\begin{enumerate}
\item{One can continue to work on the spatial manifold $R^{D-1}$
but assume the initial state is released with its super-horizon modes 
in some less singular condition \cite{AV}; or}
\item{One can work on the compact spatial manifold $T^{D-1}$
with its coordinate radius chosen such that the initial state has no 
super-horizon modes \cite{TW3}.}
\end{enumerate}
We will adopt the latter fix. Of course this makes the mode sum
discrete but the integral approximation should be excellent, and
gives a simple expression for the propagator which differs from
(\ref{modesum}) only by having an infrared cutoff at 
$k = H$.\footnote{Making the integral approximation does not
alter the renormalization of various $M_s^2 = 0$ scalar models at
one loop order \cite{vacpol,fermself,KW1,KW2}, or even at two loops
\cite{OW,BOW,PTW1,PTW2}. Because the physical graviton polarizations
have the same mode functions as scalars with $M_S^2 = 0$ 
\cite{Grishchuk}, one can also test the integral approximation with
the graviton propagator. There is no disruption of powerful
consistency checks such as the Ward identity at tree order and one 
loop \cite{TW4}, or the nature of allowed one loop counterterms 
\cite{TW5,MW1,KW3}.}

From the preceding discussion we see that the infrared corrected
propagator $i\Delta(x;z)$  is just (\ref{integral}) with the lower 
limit cutoff at $\tau = 1/\sqrt{a_x a_z}$,
\begin{eqnarray}
\lefteqn{i\Delta(x;z) = \frac{H^{D-2}}{2^D \pi^{\frac{D-3}2}} 
\int_{\frac1{\sqrt{a_x a_z}}}^{\infty} \!\!\!\!\! d\tau \, \tau^{D-2} 
\frac{J_{\frac{D-3}2}( \sqrt{a_x a_z} \, H \Delta x \tau)}{(\frac12 
\sqrt{a_x a_z} \, H \Delta x \tau)^{\frac{D-3}2} } } \nonumber \\
& & \times \Biggl\{ \theta(x^0 \!-\! z^0) H_{\nu}^{(1)}\Bigl(
\sqrt{\frac{a_z}{a_x}} \, \tau\Bigr) H_{\nu}^{(1)}\Bigl(\sqrt{\frac{a_x}{a_z}} 
\, \tau\Bigr)^* + \theta(z^0 \!-\! x^0) \Bigl({\rm conjugate}\Bigr) 
\Biggr\} . \qquad
\end{eqnarray}
Of course we can express the truncated integral as the full one
minus an integral over just the infrared,
\begin{equation}
\int_{\frac1{\sqrt{a_x a_z}}}^{\infty} \!\!\!\!\! d\tau = \int_0^{\infty} \!\!
d\tau - \int_0^{\frac1{\sqrt{a_x a_z}}} \!\!\! d\tau \quad 
\Longleftrightarrow \quad i\Delta(x;z) \equiv i\Delta^{\rm dS}(x;z) +
\Delta^{\rm IR}(x;z) \; .
\end{equation}
In this case it does not matter if dimensional regularization is used to
evaluate both $i\Delta^{\rm dS}(x;z)$ and $\Delta^{\rm IR}(x;z)$ because
the errors we make at the lower limits will cancel.

A further simplification is that $\Delta^{\rm IR}(x;z)$ only needs to include 
the infrared singular terms which grow as $a_x a_z$ increases. These terms
come entirely from the $J_{-\nu}$ parts of the Hankel function and they
are entirely real,
\begin{eqnarray}
\lefteqn{\Delta^{\rm IR}(x;z) = -\frac{H^{D-2}}{(4\pi)^{\frac{D}2}} \frac{2 
\Gamma(\nu) \Gamma(2\nu)}{\Gamma(\nu \!+\! \frac12)} 
\int_0^{\frac1{\sqrt{a_x a_z}}} \!\!\! d\tau \, \tau^{D-2} 
\frac{J_{\frac{D-3}2}( \sqrt{a_x a_z} \, H \Delta x \tau)}{(\frac12 
\sqrt{a_x a_z} \, H \Delta x \tau)^{\frac{D-3}2} } } \nonumber \\
& & \hspace{5.5cm} \times \frac{\Gamma^2(1\!-\!\nu)}{2^{2\nu}} \,
J_{-\nu}\Bigl( \sqrt{\frac{a_z}{a_x}} \, \tau\Bigr) J_{-\nu}\Bigl(
\sqrt{\frac{a_x}{a_z}} \, \tau\Bigr) \; . \qquad \label{homo}
\end{eqnarray}
Before giving the general result for $\Delta^{\rm IR}(x;z)$ it is 
instructive to work out the first two terms in the small $\tau$
expansion of the integrand,
\begin{eqnarray}
\lefteqn{\tau^{D-2} \, \frac{J_{\frac{D-3}2}( \sqrt{a_x a_z} \, H 
\Delta x \tau)}{(\frac12 \sqrt{a_x a_z} \, H \Delta x \tau)^{\frac{D-3}2} }
\frac{\Gamma^2(1 \!-\! \nu)}{2^{2\nu}} J_{-\nu}\Bigl( \sqrt{\frac{a_z}{a_x}} 
\, \tau\Bigr) J_{-\nu}\Bigl(\sqrt{\frac{a_x}{a_z}} \, \tau\Bigr) } \nonumber \\
& & \hspace{-.2cm} = \frac{\tau^{D-2-2\nu}}{\Gamma(\frac{D-1}2)} \Biggl\{1
- \frac{a_x a_z H^2 \Delta x^2 \tau^2}{2 (D \!-\! 1)} + O(\tau^4)\Biggr\}
\Biggl\{1 + \frac{(\frac{a_x}{a_z} \!+\! \frac{a_z}{a_x}) \tau^2}{4 
(\nu\!-\!1)} + O(\tau^4)\Biggr\} . \qquad
\end{eqnarray}
Now use the definition (\ref{ydef}) of the de Sitter length function 
to infer,
\begin{equation}
y = a_x a_z H^2 \Bigl[ \Delta x^2 - \Bigl(\frac1{H a_x} \!-\!
\frac1{H a_z}\Bigr)^2 \Bigr] \; \Longrightarrow \; a_x a_z H^2
\Delta x^2 = (y \!-\! 2) + \Bigl(\frac{a_x}{a_z} \!+\! \frac{a_z}{a_x}
\Bigr) \; .
\end{equation}
Hence we have,
\begin{eqnarray}
\lefteqn{\Delta^{\rm IR}(x;z) = -\frac{H^{D-2}}{(4\pi)^{\frac{D}2}} \frac{2 
\Gamma(\nu) \Gamma(2\nu)}{\Gamma(\frac{D-1}2) \Gamma(\nu \!+\! \frac12)} 
\int_0^{\frac1{\sqrt{a_x a_z}}} \!\!\! d\tau \, \tau^{D-2-2\nu} } \nonumber \\
& & \hspace{1.5cm} \times \Biggl\{1 - \Biggl[ \frac{y \!-\! 2}{2 (D\!-\!1)}
\!+\! \frac{ \nu \!-\! (\frac{D+1}2)}{2 (D \!-\!1) (\nu \!-\! 1)} \Bigl(
\frac{a_x}{a_z} \!+\! \frac{a_z}{a_x}\Bigr) \Biggr] \tau^2 + O(\tau^4) 
\Biggr\} , \qquad \\
& & = \frac{H^{D-2}}{(4\pi)^{\frac{D}2}} \frac{\Gamma(\nu) \Gamma(2\nu)}{
\Gamma(\frac{D-1}2) \Gamma(\nu \!+\! \frac12)} \Biggl\{
\frac{(a_x a_z)^{\nu - (\frac{D-1}2)}}{\nu \!-\! (\frac{D-1}2)} \nonumber \\
& & \hspace{1cm} - \Biggl[ \frac{y \!-\! 2}{2 (D\!-\!1)} \!+\! 
\frac{ \nu \!-\! (\frac{D+1}2)}{2 (D \!-\!1) (\nu \!-\! 1)} \Bigl( 
\frac{a_x}{a_z} \!+\! \frac{a_z}{a_x}\Bigr) \Biggr] \frac{(a_x a_z)^{\nu 
- (\frac{D+1}2)}}{\nu - (\frac{D+1}2)} + \dots \Biggr\} . \qquad \label{exp}
\end{eqnarray}
One can see that the first and second terms of (\ref{exp}) respectively
cancel the $N=0$ and $N=1$ divergences in the naive mode sum 
(\ref{expansion}).

To find the general form of $\Delta^{\rm IR}(x;z)$ we first note (from its
expression as an integral over $k$) that it is annihilated by $\square 
- M_S^2$. We next note from (\ref{exp}) that $\Delta^{\rm IR}(x;z)$ consists
of a series of terms, each one of which has the form $(a_x a_z)^{\nu - 
(\frac{D-1}2) - N}$ times a series involving powers of $(y-2)$ and
$(\frac{a_x}{a_z} + \frac{a_z}{a_x})$. The contributions at fixed $N$ 
must be separately annihilated by $\square - M_s^2$, and the coefficient 
of the highest power of $(y-2)$ at fixed $N$ derives enitrely from the 
$N$th order term in the expansion of the Bessel function $J_{\frac{D-3}2}$.
These two facts imply,
\begin{eqnarray}
\lefteqn{\Delta^{\rm IR}(x;z) = \frac{H^{D-2}}{(4\pi)^{\frac{D}2}} 
\frac{\Gamma(\nu) \Gamma(2\nu)}{\Gamma(\frac{D-1}2) 
\Gamma(\nu \!+\! \frac12)} } \nonumber \\
& & \hspace{.7cm} \times \sum_{N=0}^{\infty} \frac{(a_x a_z)^{\nu - 
(\frac{D-1}2) - N}}{\nu \!-\! (\frac{D-1}2) \!-\! N} \sum_{n=0}^N 
\Bigl( \frac{a_x}{a_z} \!+\! \frac{a_z}{a_x}\Bigr)^n \sum_{m=0}^{
[\frac{N-n}2]} C_{Nnm} (y \!-\!2)^{N-n-2m} \; , \qquad \label{series}
\end{eqnarray}
where the coefficients $C_{Nnm}$ are,
\begin{eqnarray}
\lefteqn{C_{Nnm} = \frac{(-\frac14)^N}{m! n! (N \!-\!n \!-\! 2m)!} \times
\frac{\Gamma(\frac{D-1}2 \!+\! N \!+\! n \!-\! \nu)}{\Gamma(\frac{D-1}2
\!+\! N \!-\! \nu)} } \nonumber \\
& & \hspace{2cm} \times \frac{\Gamma(\frac{D-1}2)}{\Gamma(\frac{D-1}2
\!+\! N\!-\! 2m)} \times \frac{\Gamma(1 \!-\!\nu)}{\Gamma(1 \!-\! \nu
\!+\! n \!+\! 2m)} \times \frac{\Gamma(1 \!-\! \nu)}{\Gamma(1 \!-\! \nu 
\!+\! m)} \; . \qquad \label{cdef}
\end{eqnarray}
Of course there is no point in extending the sum over $N$ to values
$N > \nu -\frac{D-1}2)$ for which the exponent of $a_x a_z$ becomes
negative. Those terms rapidly approach zero, and they can be dropped 
without affecting the propagator equation because they are separately
annihilated by $\square - M_s^2$.

We conclude this section by discussing three special cases which occur 
with such frequency as to merit special notation. These are 
\begin{eqnarray}
M_S^2 = (D\!-\!2) H^2 & \!\!\! \Longrightarrow \!\!\! & \nu = \frac{D\!-\!3}2 
\Rightarrow i\Delta_B(x;z) = B(y) \; , \\
M_S^2 = 0 & \!\!\! \Longrightarrow \!\!\! & \nu = \frac{D\!-\!1}2 
\Rightarrow i\Delta_A(x;z) = A(y) + \delta A(a_x,a_z,y) \; , \qquad \\
M_S^2 = -D H^2 & \!\!\! \Longrightarrow \!\!\! & \nu = \frac{D\!+\!1}2 
\Rightarrow i\Delta_W(x;z) = W(y) + \delta W(a_x,a_z,y) \; . \qquad
\end{eqnarray}
Although the $B$-type propagator is de Sitter invariant, its $A$-type
and $W$-type cousins have de Sitter breaking parts,
\begin{eqnarray}
\delta A & = & k \ln(a_x a_z) \; , \label{dsbA} \\
\delta W & = & k \Biggl\{ (D\!-\!1)^2 a_x a_z - \Bigl(\frac{D\!-\!1}2\Bigr)
\ln(a_x a_z) (y \!-\! 2) - \Bigl(\frac{a_x}{a_z} \!+\! \frac{a_z}{a_x}\Bigr)
\Biggr\} . \qquad
\end{eqnarray}
The constant $k$ is,
\begin{equation}
k \equiv \frac{H^{D-2}}{(4\pi)^{\frac{D}2}} \, \frac{\Gamma(D \!-\! 1)}{
\Gamma(\frac{D}2)} \; .
\end{equation}
The main, de Sitter invariant parts of each propagator consist of a few,
potentially ultraviolet divergent terms (at $y=0$), plus an infinite series,
\begin{eqnarray}
\lefteqn{B(y) = \frac{H^{D-2}}{(4\pi)^{\frac{D}2}} \Biggl\{
\Gamma\Bigl(\frac{D}2 \!-\!1\Bigr) \Bigl(\frac{4}{y}\Bigr)^{\frac{D}2 -1} }
\nonumber \\
& & \hspace{1.5cm} + \sum_{n=0}^{\infty} \Biggl[
\frac{\Gamma(n\!+\!\frac{D}2)}{(n \!+\! 1)!}
\Bigl(\frac{y}4 \Bigr)^{n - \frac{D}2 +2} \!\!\!\!\! -
\frac{\Gamma(n \!+\! D \!-\! 2)}{\Gamma(n \!+\! \frac{D}2)}
\Bigl(\frac{y}4 \Bigr)^n \Biggr] \Biggr\} , \qquad \label{DeltaB} \\
\lefteqn{A(y) = \frac{H^{D-2}}{(4\pi)^{\frac{D}2}} \Biggl\{
\Gamma\Bigl(\frac{D}2 \!-\!1\Bigr) \Bigl(\frac{4}{y}\Bigr)^{ 
\frac{D}2 -1} \!+\! \frac{\Gamma(\frac{D}2 \!+\! 1)}{\frac{D}2 \!-\! 2}
\Bigl(\frac{4}{y} \Bigr)^{\frac{D}2-2} \!+\! A_1 } \nonumber \\
& & \hspace{1.5cm} - \sum_{n=1}^{\infty} \Biggl[
\frac{\Gamma(n\!+\!\frac{D}2\!+\!1)}{(n\!-\!\frac{D}2\!+\!2) (n \!+\! 1)!}
\Bigl(\frac{y}4 \Bigr)^{n - \frac{D}2 +2} \!\!\!\!\! -
\frac{\Gamma(n \!+\! D \!-\! 1)}{n \Gamma(n \!+\! \frac{D}2)}
\Bigl(\frac{y}4 \Bigr)^n \Biggr] \Biggr\} , \qquad \label{DeltaA} \\
\lefteqn{W(y) = \frac{H^{D-2}}{(4\pi)^{\frac{D}2}} \Biggl\{
\Gamma\Bigl(\frac{D}2 \!-\!1\Bigr) \Bigl(\frac{4}{y}\Bigr)^{\frac{D}2 -1} 
\!+\! \frac{\Gamma(\frac{D}2 \!+\! 2)}{(\frac{D}2 \!-\! 2) (\frac{D}2 \!-\!1)}
\Bigl(\frac{4}{y} \Bigr)^{\frac{D}2-2} } \nonumber \\
& & \hspace{3cm} \!+\! \frac{\Gamma(\frac{D}2 \!+\! 3)}{2 (\frac{D}2 \!-\! 3)
(\frac{D}2\!-\!2)} \Bigl(\frac{4}{y} \Bigr)^{\frac{D}2-3} \!+\! W_1 \!+\! 
W_2 \Bigl(\frac{y \!-\!2}4\Bigr) \nonumber \\
& & \hspace{1.5cm} + \sum_{n=2}^{\infty} \Biggl[
\frac{\Gamma(n\!+\!\frac{D}2\!+\!2) (\frac{y}4)^{n-\frac{D}2+2}}{(n\!-\!
\frac{D}2\!+\!2) (n \!-\! \frac{D}2 \!+\!1) (n \!+\! 1)!} - \frac{\Gamma(n 
\!+\! D) (\frac{y}4)^n }{n (n \!-\!1) \Gamma(n \!+\! \frac{D}2)}
\Biggr] \Biggr\} . \qquad \label{DeltaW}
\end{eqnarray}
And the $D$-depdendent constants $A_1$, $W_1$ and $W_2$ are,
\begin{eqnarray}
A_1 & = & \frac{\Gamma(D\!-\!1)}{\Gamma(\frac{D}2)} \Biggl\{ 
-\psi\Bigl(1 \!-\! \frac{D}2\Bigr) + \psi\Bigl(\frac{D\!-\!1}2\Bigr) +
\psi(D \!-\!1) + \psi(1) \Biggr\} , \\
W_1 & = & \frac{\Gamma(D\!+\!1)}{\Gamma(\frac{D}2 \!+\!1)} \Biggl\{ 
\frac{D \!+\!1}{2 D} \Biggr\} , \\
W_2 & = & \frac{\Gamma(D\!+\!1)}{\Gamma(\frac{D}2 \!+\!1)} \Biggl\{ 
\psi\Bigl(-\frac{D}2\Bigr) - \psi\Bigl(\frac{D\!+\!1}2\Bigr) -
\psi(D \!+\!1) - \psi(1) \Biggr\} .
\end{eqnarray}

The infinite series terms of $B(y)$, $A(y)$ and $W(y)$ makes expressions
(\ref{DeltaB}-\ref{DeltaW}) seem intimidating. However, note that each
pair of terms in the infinite sums cancels for $D=4$, so they only need 
to be retained when multiplying a potentially divergent quantity. Further,
because $y^n$ vanishes more and more strongly at coincidence as $n$ 
increases, only a handful of the smallest $n$ terms ever need to be
included. This makes loop computations manageable. For a massless, 
minimally coupled scalar with a quartic self-interaction, two loop 
results have been obtained for the expectation value of the stress tensor 
\cite{OW}, for the scalar self-mass-squared \cite{BOW} and for the 
quantum-corrected mode functions \cite{KO}. In Yukawa theory it has been 
used to compute the expectation value of the coincident vertex function 
at two loop order \cite{MW2}, and it has been used for a variety of two 
loop computations in scalar quantum electrodynamics \cite{PTW1,PTW2}. 

The need for de Sitter breaking terms in $i\Delta_A(x;z)$ has long
been recognized \cite{AF}, and ours reproduces the classic and well 
known result for the coincidence limit of the propagator \cite{classic}.
The de Sitter breaking terms also show up in the differential equations
obeyed by the de Sitter invariant parts of the various propagators,
\begin{eqnarray}
(4 y \!-\! y^2) B'' + D (2 \!-\! y) B' - (D\!-\!2) B & = & 0 \; ,\label{Beqn}\\
(4 y \!-\! y^2) A'' + D (2 \!-\! y) A' & = & (D \!-\! 1) k \; , \\
(4 y \!-\! y^2) W'' + D (2 \!-\! y) W' + D W & = & \frac12 (D \!+\!1)
(D\!-\!1) k (2 \!-\! y) \; .
\end{eqnarray}
Whereas the equation for $B(y)$ is homogeneous, the equations for
$A(y)$ and $W(y)$ both possess inhomgeneous terms which are cancelled
by $\square - M_S^2$ acting on the de Sitter breaking terms 
$\delta A(a_x,a_z,y)$ and $\delta W(a_x,a_z,y)$. Finally, we give some
differential relations between the de Sitter invariant parts which 
follow from the series expansions (\ref{DeltaB}-\ref{DeltaW}) \cite{MTW},
\begin{eqnarray}
(2 \!-\! y) A' - k & = & 2 B' \; , \label{rel1} \\
(2 \!-\! y) W'' +\frac12 (D \!-\!1) k & = & 2 A'' \; .
\end{eqnarray}

\section{Vectors}

The purpose of this section is to demonstrate that the de Sitter invariant
propagator equation possesses no de Sitter invariant solution for a massive
vector with $M_V^2 \leq -2(D-1) H^2$. We begin by explaining how the full 
vector propagator (longitudinal plus transverse) can be written as the 
transverse vector propagator plus a double gradient of the difference of 
two known scalar propagators. We then derive a formal, de Sitter invariant 
solution for the transverse vector propagator in terms of scalar propagators. 
One of these scalar propagators possesses the infrared divergences we found 
in the previous section. We show how the problem can be corrected, and we 
derive the leading infrared correction. 

The full propagator for a massive vector obeys the equation,
\begin{equation}
\sqrt{-g(x)} \, \Bigl[ \square_x - (D\!-\!1) H^2 - M_V^2\Bigr] \,
i\Bigl[\mbox{}_{\mu} \Delta_{\nu}\Bigr](x;z) = g_{\mu\nu} \, i
\delta^D(x \!-\!z) \; . \label{FPeqn}
\end{equation}
Although this object does appear in certain projection operators, there
is greater physical interest in its transverse part which obeys,
\begin{equation}
g^{\mu\nu}(x) \frac{D}{D x^{\mu}} i\Bigl[\mbox{}_{\nu} \Delta^{\rm T}_{\rho}
\Bigr](x;z) = 0 = g^{\rho\sigma}(z) \frac{D}{D z^{\rho}} i\Bigl[\mbox{}_{\mu} 
\Delta^{\rm T}_{\sigma}\Bigr](x;z) \; . \label{transcon}
\end{equation}
An excellent early study of the massive, transverse vector propagator was 
carried out by Allen and Jacobson \cite{AJ}. A minor error in their work 
is that the source term is not transverse. When this is corrected the 
propagator equation reads \cite{TW6},
\begin{eqnarray}
\lefteqn{\sqrt{-g(x)} \Bigl[\square_x - (D\!-\!1) H^2 - M_V^2\Bigr] 
i\Bigl[\mbox{}_{\mu} \Delta^{\rm T}_{\nu}\Bigr](x;z) } \nonumber \\
& & \hspace{4cm} = g_{\mu\nu} i\delta^D(x \!-\! z) + \sqrt{-g(x)} \,
\frac{\partial}{\partial x^{\mu}} \frac{\partial}{\partial z^{\nu}} 
i\Delta_A(x;z) \; . \qquad \label{VPeqn}
\end{eqnarray}
Although $i\Delta_A(x;z)$ contains a de Sitter breaking part,
$k \ln(a_x a_z)$, this makes no contribution when the propagator
is differentiated on both of its arguments. Therefore, equation 
(\ref{VPeqn}) is fully de Sitter invariant.

Given the transverse vector propagator one can construct the full vector 
propagator by adding the double gradient of a longitudinal part,
\begin{equation}
i\Bigl[\mbox{}_{\mu} \Delta_{\nu}\Bigr](x;z) =
i\Bigl[\mbox{}_{\mu} \Delta^{\rm T}_{\nu}\Bigr](x;z) +
\frac{\partial}{\partial x^{\mu}} \frac{\partial}{\partial z^{\nu}}
\, i\Delta^{\rm L}(x;z) \; . \label{fullprop}
\end{equation}
The equation obeyed by $i\Delta^{\rm L}(x;z)$ follows from substituting
(\ref{fullprop}) in (\ref{FPeqn}), commuting some derivatives and using 
(\ref{VPeqn}) to conclude,
\begin{equation}
\frac{\partial}{\partial x^{\mu}} \frac{\partial}{\partial z^{\nu}} 
\Biggl\{ i\Delta_A(x;z) + \Bigl[ \square_x - M_V^2\Bigr] \,
i \Delta^{\rm L}(x;z) \Biggr\} = 0 . \label{Leqn}
\end{equation}
A solution to (\ref{Leqn}) is \cite{MTW},
\begin{equation}
i\Delta^{\rm L}(x;z) = \frac1{M_V^2} \Bigl[ i\Delta_A(x;z) - i\Delta_M(x;z)
\Bigr] \; , \label{DeltaL}
\end{equation}
where $i\Delta_M(x;z)$ is the scalar propagator with mass $M_S^2 = M_V^2$.

Because we can always construct the full vector propagator from its
transverse part by the procedure just described, we will henceforth
concentrate on the transverse vector propagator. The most general de 
Sitter invariant, symmetric bi-tensor which obeys the transversality 
condition (\ref{transcon}) can be expressed using an arbitrary function 
$\gamma(y)$ and the basis tensors of section 2 \cite{TW6},
\begin{eqnarray}
\lefteqn{ i\Bigl[\mbox{}_{\mu} \Delta^{\rm dS}_{\nu}\Bigr](x;z) =
\Biggl[\frac{-(4y \!-\! y^2) \gamma' \!-\! (D\!-\!1) (2\!-\!y) \gamma}{
4 (D\!-\!1) H^2} \Biggr] \frac{\partial^2 y(x;z)}{\partial x^{\mu} 
\partial z^{\nu}} } \nonumber \\
& & \hspace{6cm} + \Biggl[ \frac{(2 \!-\!y) \gamma' \!-\! (D\!-\!1) \gamma}{
4 (D\!-\!1) H^2} \Biggr] \frac{\partial y}{\partial x^{\mu}}
\frac{\partial y}{\partial z^{\nu}} \; . \qquad \label{dSprop}
\end{eqnarray}
Substituting (\ref{dSprop}) into (\ref{VPeqn}) gives the following
differential equation for $\gamma(y)$ away from coincidence (that is,
away from $x^{\mu} = z^{\mu}$) \cite{TW6},
\begin{equation}
(4y \!-\! y^2) \gamma'' + (D\!+\!2) (2\!-\!y) \gamma' - 
\Bigl[2 (D\!-\!1) + \frac{M_V^2}{H^2}\Bigr] \gamma
= 2 (D\!-\!1) B'(y) \; . \qquad \label{gameqn}
\end{equation}
Recovering the delta function in (\ref{VPeqn}) additionally requires that
the most singular term for $y \rightarrow 0$ must be \cite{TW6},
\begin{equation}
\gamma(y) = \frac{H^{D-2}}{(4\pi)^{\frac{D}2}} \Biggl\{ 
\Bigl(\frac{D \!-\! 1}2\Bigr) \Gamma\Bigl(\frac{D}2 \!-\!1\Bigr) 
\Bigl(\frac{4}{y}\Bigr)^{\frac{D}2-1} + O(y^{2-\frac{D}2}) \Biggr\} . 
\label{sing}
\end{equation}

At this point it helps to note that the left hand side of relation 
(\ref{gameqn}) can be expressed as the derivative of a scalar kinetic 
operator,
\begin{eqnarray}
\lefteqn{(4y \!-\! y^2) \gamma'' + (D\!+\!2) (2\!-\!y) \gamma' - 
\Bigl[2 (D\!-\!1) + \frac{M_V^2}{H^2}\Bigr] \gamma } \nonumber \\
& & \hspace{5cm} = \frac1{H^2} \frac{\partial}{\partial y} \Bigl[
\square_x - (D\!-\!2) H^2 - M_V^2\Bigr] I[\gamma] \; , \qquad
\end{eqnarray}
where $I[y]$ stands for the indefinite integral of $\gamma(y)$ with
respect to $y$. Hence $I[\gamma]$ obeys the scalar propagator equation,
\begin{equation}
\Bigl[\square - (D\!-\!2) H^2 - M_V^2\Bigr] I[\gamma] = 
2 (D\!-\!1) H^2 B \; .
\end{equation}
This is very like the equation we just solved for the longitudinal
part of the vector propagator so it should not seem surprising that
the unique solution for relations (\ref{gameqn}-\ref{sing}) is,
\begin{equation}
\gamma(y) = 2 (D\!-\!1) \frac{H^2}{M_V^2} \frac{\partial}{\partial y}
\Bigl[ E(y) - B(y)\Bigr] \; , \label{ansatz}
\end{equation}
where $B(y)$ is the scalar propagator with mass $M_S^2 = (D-2) H^2$ and
$E(y)$ is the scalar propagator with mass $M_S^2 = (D-2) H^2 + M_V^2$.
To verify (\ref{ansatz}), first note the two leading terms in the series 
expansions of $B(y)$ and $E(y)$, from expressions (\ref{DeltaB}) and 
(\ref{expansion}) with $\nu^2 = (\frac{D-3}2)^2 - \frac{M_V^2}{H^2}$, 
\begin{eqnarray}
B(y) & \!\!=\!\! & \frac{H^{D-2}}{(4\pi)^{\frac{D}2}} \Biggl\{ \Gamma(\frac{D}2
\!-\! 1\Bigr) \Bigl(\frac{4}{y}\Bigr)^{\frac{D}2-1} + \Gamma\Bigl(\frac{D}2
\Bigr) \Bigl(\frac{4}{y}\Bigr)^{\frac{D}2-2} + \dots \Biggr\} , \\
E(y) & \!\!=\!\! & \frac{H^{D-2}}{(4\pi)^{\frac{D}2}} \Biggl\{ \Gamma(\frac{D}2
\!-\! 1\Bigr) \Bigl(\frac{4}{y}\Bigr)^{\frac{D}2-1} + \Bigl(\nu^2 \!-\!
\frac14\Bigr) \Gamma\Bigl(\frac{D}2 \!-\! 2\Bigr) \Bigl(\frac{4}{y}
\Bigr)^{\frac{D}2-2} + \dots \Biggr\} . \qquad
\end{eqnarray}
This establishes that (\ref{ansatz}) has the correct singularity 
(\ref{sing}) near coincidence. To check the off-coincidence condition 
(\ref{gameqn}), note that $B(y)$ obeys (\ref{Beqn}) and $E(y)$ obeys,
\begin{equation}
(4 y \!-\! y^2) E'' + D (2 \!-\! y) E' - \Bigl[(D \!-\! 2) \!+\!
\frac{M_V^2}{H^2}\Bigr] E = 0 \; . \label{Eeqn}
\end{equation}
Differentiating (\ref{Beqn}) and (\ref{Eeqn}) with respect to $y$ gives,
\begin{eqnarray}
(4 y \!-\! y^2) B''' + (D \!+\! 2) (2 \!-\! y) B'' - \Bigl[2 (D \!-\! 1) \!+\!
\frac{M_V^2}{H^2}\Bigr] B' & = & -\frac{M_V^2}{H^2} \, B' \; , \qquad
\label{G1} \\
(4 y \!-\! y^2) E''' + (D \!+\! 2) (2 \!-\! y) E'' - \Bigl[2 (D \!-\! 1) \!+\!
\frac{M_V^2}{H^2}\Bigr] E' & = & 0 \; . \label{G2}
\end{eqnarray}
Subtracting (\ref{G1}) from (\ref{G2}) and multiplying by $2(D-1)
\frac{H^2}{M_V^2}$ shows that our ansatz (\ref{ansatz}) indeed obeys 
equation (\ref{gameqn}).

Of course using our ansatz (\ref{ansatz}) in (\ref{dSprop}) only makes
sense if the two scalar propagators $B(y)$ and $E(y)$ exist! Recall
that they correspond to masses $M_S^2 = (D-2)H^2$ and $M_S^2 = (D-2)H^2
+ M_V^2$, respectively. For $D > 2$ the $B$-type propagator is safe but
the $E$-type propagator can have problems if $M_V^2$ is sufficiently
negative. Now consider its formal series expansion (\ref{expansion})
with $\nu^2 = (\frac{D-3}2)^2 - \frac{M_V^2}{H^2}$, and in particular,
the problematic gamma function, $\Gamma(\frac{D-1}2 \!-\! \nu \!+\! n)$. 
Because it is really the derivative $E'(y)$ which appears in our ansatz 
(\ref{ansatz}), the problem at $n=0$ drops out, but there are divergences
at $n = N+1$ for all non-negative integers $N$. This corresponds to the
following vector masses,
\begin{equation}
\nu = \Bigl(\frac{D-1}2\Bigr) + N + 1 \qquad \Longleftrightarrow \qquad
M_V^2 = - (N \!+\! 2) (N \!+\! D \!-\! 1) H^2 \; . \label{Vlogs}
\end{equation}
As emphasized in the previous section, condition (\ref{Vlogs}) only
gives the {\it logarithmic} infrared divergences. For $M_V^2 \leq -2
(D-1) H^2$ there will be power law infrared divergences which analytic
continuation techniques incorrectly subtract off. We turn now to 
fixing this problem.

Because the infrared divergences we have found for $M_V^2 \leq -2(D-1) H^2$ 
arise from assuming a de Sitter invariant solution to the propagator 
equation (\ref{VPeqn}), they can only be corrected by abandoning that 
assumption. However, there is no need to discard the formal de Sitter
invariant solution $i[\mbox{}_{\mu}\Delta^{\rm dS}_{\nu}](x;z)$. Just
like the scalar of the preceding section, we need only to add to it a 
de Sitter breaking, infrared correction,
\begin{equation}
i \Bigl[\mbox{}_{\mu} \Delta^{\rm T}_{\nu}\Bigr](x;z) \equiv 
i \Bigl[\mbox{}_{\mu} \Delta^{\rm dS}_{\nu}\Bigr](x;z) +
\Bigl[\mbox{}_{\mu} \Delta^{\rm IR}_{\nu}\Bigr](x;z) \; .
\end{equation}
Of course the infrared correction must be symmetric and it must obey 
the transversality condition (\ref{transcon}). In analogy with the
scalar, we also want it to be annihilated by the kinetic operator
$[\square - (D\!-\!1) H^2 - M_V^2]$.

Abandoning de Sitter invariance in $[\mbox{}_{\mu} \Delta^{\rm 
IR}_{\nu}](x;z)$ affects the tensor structure as well as the scalar 
coefficient functions. If we preserve spatial homogeneity and isotropy 
(as we did for the scalars of the previous section) then the only extra 
basis tensors we require are the derivatives of $u(x;z) \equiv 
\ln(a_x a_z)$ with respect to $x^{\mu}$ and $z^{\nu}$,
\begin{equation}
\frac{\partial u}{\partial x^{\mu}} = H a_x \delta^0_{\mu} \qquad , \qquad
\frac{\partial u}{\partial z^{\nu}} = H a_z \delta^0_{\nu} \; .
\end{equation}
The most general homogeneous and isotropic tensor with the correct
symmetries is,
\begin{eqnarray}
\lefteqn{ \Bigl[\mbox{}_{\mu} \Delta^{\rm IR}_{\nu}\Bigr](x;z) = 
F_1(a_x,a_z,y) \frac{\partial^2 y}{\partial x^{\mu} \partial z^{\nu}} 
+ F_2(a_x,a_z,y) \frac{\partial y}{\partial x^{\mu}} 
\frac{\partial y}{\partial z^{\nu}} } \nonumber \\
& & \hspace{-.3cm} + F_3(a_x,a_z,y) \frac{\partial u}{\partial x^{\mu}}
\frac{\partial y}{\partial z^{\nu}} + F_3(a_z,a_x,y) 
\frac{\partial y}{\partial x^{\mu}} \frac{\partial u}{\partial z^{\nu}} +
F_4(a_x,a_z,y) \frac{\partial u}{\partial x^{\mu}} 
\frac{\partial u}{\partial z^{\nu}} \; . \qquad \label{vecans}
\end{eqnarray}
The coefficient functions $F_1$, $F_2$ and $F_4$ must be symmetric under
interchange of $a_x$ and $a_z$. $F_3$ does not have this symmetry; when
the argument lists are suppressed (to save space) we will indicate 
the interchange of $a_x$ and $a_z$ by a superscript $T$,
\begin{equation}
F_3^T(a_x,a_z,y) \equiv F_3(a_z,a_x,y) \; .
\end{equation}
We will also use a prime to denote differentiation with respect to $y$,
\begin{equation}
F_i'(a_x,a_z,y) \equiv \frac{\partial}{\partial y} \, F_i(a_x,a_z,y) \; .
\end{equation}
The right way to think about the coefficient functions $F_i(a_x,a_z,y)$
is that enforcing transversality determines $F_3$ and $F_4$ in terms of 
$F_1$ and $F_2$. Then $F_1$ and $F_2$ are fixed (up to normalization of
each independent term) by demanding that $[\square - (D\!-\!1) H^2 - M_V^2]$ 
annihilates $[\mbox{}_{\mu} \Delta^{\rm IR}_{\nu}](x;z)$.

Covariant derivatives of the new tensor involve some extra identities in
addition to those of section 2,
\begin{equation}
\frac{D^2 u}{D x^{\mu} D x^{\nu}} = -H^2 g_{\mu\nu}(x) - 
\frac{\partial u}{\partial x^{\mu}} \frac{\partial u}{\partial x^{\nu}}
\quad , \quad \frac{D^2 u}{D z^{\mu} D z^{\nu}} = -H^2 g_{\mu\nu}(z) - 
\frac{\partial u}{\partial z^{\mu}} \frac{\partial u}{\partial z^{\nu}} \; .
\end{equation}
There are also some new contraction identities,
\begin{eqnarray}
g^{\mu\nu}(x) \frac{\partial u}{\partial x^{\mu}}
\frac{\partial u}{\partial x^{\nu}} & = & - H^2 \; , \\
g^{\mu\nu}(x) \frac{\partial u}{\partial x^{\mu}}
\frac{\partial y}{\partial x^{\nu}} & = & - H^2 \Bigl[ y \!-\! 2 
+ 2 \frac{a_z}{a_x} \Bigr] \; , \\
g^{\mu\nu}(x) \frac{\partial u}{\partial x^{\mu}}
\frac{\partial^2 y}{\partial x^{\nu} \partial z^{\rho}} & = & - H^2 
\Bigl[ \frac{\partial y}{\partial z^{\rho}} + 2 \frac{a_z}{a_x} 
\frac{\partial u}{\partial z^{\rho}} \Bigr] \; .
\end{eqnarray}
The covariant divergence $g^{\mu\nu}(x) D_{\mu} i[\mbox{}_{\nu} 
\Delta^{\rm IR}_{\rho}](x;z)$ produces one term proportional to 
$\partial y/\partial z^{\rho}$ and another proportional to
$\partial u/\partial z^{\rho}$. The term proportional to
$\partial y/\partial z^{\rho}$ gives an equation for the
coefficient function $F_3(a_x,a_z,y)$ in terms of $F_1(a_x,a_z,y)$
and $F_2(a_x,a_z,y)$,
\begin{eqnarray}
\lefteqn{-(2\!-\!y) F_3' + D F_3 + a_x \frac{\partial F_3}{\partial a_x}
+ 2 \frac{a_z}{a_x} F_3' = (2 \!-\!y) F_1' - D F_1
- a_x \frac{\partial F_1}{\partial a_x} } \nonumber \\
& & \hspace{1.7cm} + (4y \!-\!y^2) F_2' + (D\!+\!1) (2 \!-\! y) F_2
+ (2 \!-\!y) a_x \frac{\partial F_2}{\partial a_x} - 2 a_z
\frac{\partial F_2}{\partial a_x} \; . \qquad \label{F3}
\end{eqnarray}
Because the equation involves only first derivatives with respect to
$y$ and $a_x$, the general solution can be given using the Method of
Characteristics. Once $F_3(a_x,a_z,y)$ is known, the transversality
relation proportional to $\partial u/\partial z^{\rho}$ gives an
equation for the remaining coefficient function $F_4(a_x,a_z,y)$,
\begin{eqnarray}
\lefteqn{-(2\!-\!y) F_4' + (D\!-\!1) F_4+ a_x \frac{\partial F_4}{\partial a_x}
+ 2 \frac{a_z}{a_x} F_4' = -2 a_z \frac{\partial F_1}{\partial a_x} -
2 \frac{a_z}{a_x} F_3 } \nonumber \\
& & \hspace{2cm} + (4y \!-\!y^2) {F_3^T}' + D (2 \!-\! y) F_3^T
+ (2 \!-\!y) a_x \frac{\partial F_3^T}{\partial a_x} -
2 a_z \frac{\partial F_3^T}{\partial a_x} \; . \qquad \label{F4}
\end{eqnarray}
This equation can also be solved by the Method of Characteristics.

Once one has determined the coefficient functions $F_3$ and $F_4$ from
equations (\ref{F3}-\ref{F4}), $F_1$ and $F_2$ are obtained, up to 
normalization of each independent piece, by requiring 
$[\mbox{}_{\mu} \Delta^{\rm IR}_{\nu}](x;z)$ to solve the homogeneous 
propgator equation,
\begin{equation}
\Bigl[ \square_x - (D\!-\!1) H^2 - M_V^2\Bigr] \, \Bigl[\mbox{}_{\mu}
\Delta^{\rm IR}_{\nu}\Bigr](x;z) = 0 \; . 
\end{equation}
Of course this gives a relation for each of the five tensor structures
present in $[\mbox{}_{\mu} \Delta^{\rm IR}_{\nu}](x;z)$, however, only
two of these five relations are independent. The one proportional to
$\partial^2 y/\partial x^{\mu} \partial z^{\nu}$ constrains 
$F_1(a_x,a_z,y)$,
\begin{eqnarray}
\lefteqn{(4y \!-\! y^2) F_1'' + D (2 \!-\!y) F_1' - \Bigl[D \!+\!
\frac{M_V^2}{H^2} \Bigr] F_1 + 2 (2 \!-\!y) F_2  + 2(2 \!-\!y) a_x
\frac{\partial F_1'}{\partial a_x} } \nonumber \\
& & \hspace{2cm} - 4 a_z \frac{\partial F_1'}{\partial a_x} - (D\!-\!1) 
a_x \frac{\partial F_1}{\partial a_x} - a_x \frac{\partial}{\partial a_x} 
\Bigl[ a_x \frac{\partial F_1}{\partial a_x} \Bigr] - 2 F_3 = 0 \; . 
\qquad \label{F1}
\end{eqnarray}
And $F_2(a_x,a_z,y)$ is constrained by the term proportional to
$\partial y/\partial x^{\mu} \, \partial y/\partial z^{\nu}$,
\begin{eqnarray}
\lefteqn{(4y \!-\! y^2) F_2'' + (D\!+\!4) (2 \!-\!y) F_2' - \Bigl[2D \!+\!
\frac{M_V^2}{H^2} \Bigr] F_2 - 2 F_1' + 2(2 \!-\!y) a_x
\frac{\partial F_2'}{\partial a_x} } \nonumber \\
& & \hspace{2cm} - 4 a_z \frac{\partial F_2'}{\partial a_x} - (D\!+\!1) 
a_x \frac{\partial F_2}{\partial a_x} - a_x \frac{\partial}{\partial a_x} 
\Bigl[ a_x \frac{\partial F_2}{\partial a_x} \Bigr] - 2 F_3' = 0 \; . 
\qquad \label{F2}
\end{eqnarray}

The normalization comes from differentiating (\ref{F1}) with respect to 
$y$ and then subtracting (\ref{F2}) to obtain an equation for the 
combination $F_1' - F_2$. By making the definition,
\begin{equation}
F_1'(a_x,a_z,y) - F_2(a_x,a_z,y) \equiv \mathcal{E}'(a_x,a_z,y) \; ,
\end{equation}
we can identify this relation as the homogeneous equation for an
$E$-type scalar with mass $M_s^2 = (D\!-\!2) H^2 + M_V^2$,
\begin{equation}
\frac{\partial}{\partial y} \, (\ref{F1}) - (\ref{F2}) = H^{-2}
\frac{\partial}{\partial y} \Bigl[\square_x - (D\!-\!2)H^2 - M_V^2\Bigr]
\, \mathcal{E} = 0 \; .
\end{equation}
Now write the de Sitter invariant part of the propagator in a form
similar to the de Sitter breaking part,
\begin{equation}
i\Bigl[\mbox{}_{\mu} \Delta^{\rm dS}_{\nu}\Bigr](x;z) \equiv 
\mathcal{F}_1(y) \frac{\partial^2 y}{\partial x^{\mu} \partial z^{\nu}}
+ \mathcal{F}_2(y) \frac{\partial y}{\partial x^{\mu}} \frac{\partial y}{
\partial z^{\nu}} \; .
\end{equation}
From relations (\ref{dSprop}-\ref{gameqn}) and (\ref{ansatz}) we infer,
\begin{equation}
\mathcal{F}_1' - \mathcal{F}_2 = -\Biggl[ \frac{(4 y \!-\! y^2) \gamma''
\!+\! (D \!+\!2) (2 \!-\!y) \gamma' \!-\! 2(D\!-\!1) \gamma}{4 (D\!-\!1) H^2}
\Biggr] = -\frac{ E'(y)}{2 H^2} \; .
\end{equation}
Hence we infer,
\begin{eqnarray}
\lefteqn{F_1'(a_x,a_z,y) - F_2(a_x,a_z,y) = -\frac1{2 H^2} 
\frac{\partial}{\partial y} \, \Delta^{\rm IR}(x;z) \; , } \\
& & = -\frac1{2 H^2} \, \frac{H^{D-2}}{(4\pi)^{\frac{D}2}} 
\frac{\Gamma(\nu) \Gamma(2\nu)}{\Gamma(\frac{D-1}2) 
\Gamma(\nu \!+\! \frac12)} \sum_{N=0}^{\infty} \frac{(a_x a_z)^{\nu - 
(\frac{D+1}2) - N}}{\nu \!-\! (\frac{D+1}2) \!-\! N} \nonumber \\
& & \hspace{.9cm} \times \sum_{n=0}^{N+1} \Bigl( \frac{a_x}{a_z} \!+\! 
\frac{a_z}{a_x}\Bigr)^n \sum_{m=0}^{[\frac{N+1-n}2]} (N\!+\!1\!-\!n \!-\! 2m) 
C_{N+1 \, nm} (y \!-\!2)^{N-n-2m} , \qquad \label{normalization}
\end{eqnarray}
where $\Delta^{\rm IR}(x;z)$ was defined in equation (\ref{series}), the
coefficients $C_{Nnm}$ are given in (\ref{cdef}), and the index $\nu$ obeys 
$\nu^2 = (\frac{D-3}2)^2 - \frac{M_V^2}{H^2}$.

It is instructive to give the first ($N=0$) term in the series expansions
of the four coefficient functions $F_i(a_x,a_z,y)$,
\begin{eqnarray}
(F_1)_{0} & \!=\! & \mathcal{F} \Biggl[ (y \!-\! 2) + \frac{(\nu \!-\! 
\frac{D+1}2) (\nu \!+\! \frac{D-1}2)}{(\nu \!+\! \frac{D-3}2) (\nu \!-\! 1)} 
\Bigl(\frac{a_x}{a_z} \!+\! \frac{a_z}{a_x}\Bigr) \Biggr] 
\frac{(a_x a_z)^{\nu - \frac{D+1}2}}{\nu \!-\! \frac{D+1}2} \; , \qquad \\
(F_2)_{0} & \!=\! & -\mathcal{F} \frac{(a_x a_z)^{\nu - \frac{D+1}2}}{
\nu \!-\! \frac{D+1}2} \; , \qquad \\
(F_3)_{0} & \!=\! & -\mathcal{F} \frac{(\nu \!-\! \frac{D+1}2)}{(\nu \!+\!
\frac{D-3}2)} \Biggl[ \frac{(\frac{D+1}2)}{(\nu \!-\! 1)} \Bigl(
\frac{a_x}{a_z} \!+\! \frac{a_z}{a_x}\Bigr) + \Bigl(\frac{a_x}{a_z} \!-\!
\frac{a_z}{a_x}\Bigr)\Biggr] \frac{(a_x a_z)^{\nu - \frac{D+1}2}}{\nu \!-\! 
\frac{D+1}2} \; , \qquad \\
(F_4)_{0} & \!=\! & -\mathcal{F} \frac{(\nu \!-\! \frac{D+1}2)}{(\nu \!+\!
\frac{D-3}2)} \Biggl[ \frac{(\nu \!-\! \frac{D+3}2)}{(\nu \!-\! 1)} \Bigl(
\frac{a_x}{a_z} \!+\! \frac{a_z}{a_x}\Bigr) (y \!-\! 2) \nonumber \\
& & \hspace{5.8cm} + 4 \frac{(\nu \!-\! \frac{D+1}2)}{(\nu \!+\! \frac{D-3}2)} 
\Biggr] \frac{(a_x a_z)^{\nu-\frac{D+1}2}}{\nu \!-\! \frac{D+1}2} \; . \qquad
\end{eqnarray}
Here the index $\nu$ obeys $\nu^2 = (\frac{D-3}2)^2 - \frac{M_V^2}{H^2}$ and
the constant $\mathcal{F}$ is,
\begin{equation}
\mathcal{F} \equiv \frac1{16 H^2} \, \frac{H^{D-2}}{(4\pi)^{\frac{D}2}} \,
\frac{\Gamma(\nu) \Gamma(2\nu)}{\Gamma(\frac{D+1}2) \Gamma(\nu \!+\! \frac12)}
\; . \label{Fmath}
\end{equation}
The general series expansion for $F_1(a_x,a_z,y)$ has the form,
\begin{eqnarray}
\lefteqn{F_1 = \mathcal{F} \sum_{N=0}^{\infty} \frac{(a_x a_z)^{\nu-
(\frac{D+1}2)-N}}{\nu \!-\! (\frac{D+1}2) \!-\! N} } \nonumber \\
& & \hspace{3.5cm} \times
\sum_{n=0}^{N+1} \Bigl( \frac{a_x}{a_z} \!+\! \frac{a_z}{a_x}\Bigr)^n 
\sum_{m=0}^{[\frac{N+1-n}2]} F^1_{Nnm} (y \!-\!2)^{N+1-n-2m} \; . 
\qquad \label{F1exp}
\end{eqnarray}
As with the de Sitter breaking corrections for the scalar, there is no
point in extending the series for $N > \nu - \frac{D+1}2$. The other 
coefficient functions have the same first line as (\ref{F1exp}) so we 
give only their subsequent forms,
\begin{eqnarray}
F_2 & \longrightarrow & 
\sum_{n=0}^{N} \Bigl( \frac{a_x}{a_z} \!+\! \frac{a_z}{a_x}\Bigr)^n 
\sum_{m=0}^{[\frac{N-n}2]} F^2_{Nnm} (y \!-\!2)^{N-n-2m} \; , \qquad \\
F_3 & \longrightarrow & 
\sum_{n=0}^{N} \Bigl( \frac{a_x}{a_z} \!+\! \frac{a_z}{a_x}\Bigr)^n 
\nonumber \\
& & \hspace{.5cm} \times \sum_{m=0}^{[\frac{N-n}2]} \Biggl[
F^{3a}_{Nnm} \Bigl(\frac{a_x}{a_z} \!+\!  \frac{a_z}{a_x}\Bigr) +
F^{3b}_{Nnm} \Bigl(\frac{a_x}{a_z} \!-\!  \frac{a_z}{a_x}\Bigr) \Biggr]
(y \!-\!2)^{N-n-2m} \; , \qquad \\
F_4 & \longrightarrow & 
\sum_{n=0}^{N+1} \Bigl( \frac{a_x}{a_z} \!+\! \frac{a_z}{a_x}\Bigr)^n 
\sum_{m=0}^{[\frac{N+2-n}2]} F^4_{Nnm} (y \!-\!2)^{N+2-n-2m} \; . \qquad 
\label{F4exp}
\end{eqnarray}

We close this section by giving the transverse vector propagator for
the important special case of, $M_V^2 = -2(D\!-\!1) H^2$. For that mass
the associated scalar has $M_S^2 = -D H^2$, corresponding to the $W$-type
propagator considered at the end of section 3. For this reason we give
the vector propagator a subscript $W$, and we decompose it into a de
Sitter invariant part and a de Sitter breaking part,
\begin{equation}
i\Bigl[\mbox{}_{\mu} \Delta^{\rm T}_{\nu}\Bigr]_{W}(x;z) =
\Bigl[\mbox{}_{\mu} W_{\nu}\Bigr](x;z) +
\Bigl[\mbox{}_{\mu} \delta W_{\nu}\Bigr](x;z) \; .
\end{equation}
The de Sitter invariant part is,
\begin{eqnarray}
\lefteqn{\Bigl[\mbox{}_{\mu} W_{\nu}\Bigr](x;z) = \Biggl[ \frac{ (4y\!-\!y^2) 
(W\!-\!B)'' \!+\! (D\!-\!1) (2 \!-\! y) (W\!-\!B)'}{4 (D\!-\!1) H^2} \Biggr]
\frac{\partial^2 y}{\partial x^{\mu} \partial z^{\nu}} } \nonumber \\
& & \hspace{3cm} + \Biggl[ \frac{ -(2 \!-\! y) (W\!-\!B)'' \!+\! (D\!-\!1) 
(W\!-\!B)'}{4 (D\!-\!1) H^2} \Biggr] \frac{\partial y}{\partial x^{\mu}} 
\frac{\partial y}{\partial z^{\nu}} \; . \qquad
\end{eqnarray}
And the de Sitter breaking part is,
\begin{eqnarray}
\lefteqn{\Bigl[\mbox{}_{\mu} \delta W_{\nu}\Bigr] = \frac{k}{4 (D\!-\!1)
H^2} \Biggl\{\Biggl[\frac{(D\!-\!1)^2}2 \ln(a_x a_z) (y \!-\!2) \!+\! D \Bigl(
\frac{a_x}{a_z} \!+\! \frac{a_z}{a_x}\Bigr)\Biggr]
\frac{\partial^2 y}{\partial x^{\mu} \partial z^{\nu}} } \nonumber \\
& & \hspace{2cm} - \frac{(D\!-\!1)^2}2 \ln(a_x a_z) \frac{\partial y}{
\partial x^{\mu}} \frac{\partial y}{\partial z^{\nu}} \!-\! \Biggl[ D
\frac{a_x}{a_z} \!+\! \frac{a_z}{a_x} \Biggr] \frac{\partial u}{\partial 
x^{\mu}} \frac{\partial y}{\partial z^{\nu}} \nonumber \\
& & \hspace{3cm} - \Biggl[ \frac{a_x}{a_z} \!+\! D \frac{a_z}{a_x} \Biggr]
\frac{\partial y}{\partial x^{\mu}} \frac{\partial u}{\partial z^{\nu}} \!+\! 
\Bigl(\frac{a_x}{a_z} \!+\! \frac{a_z}{a_x}\Bigr) (y \!-\! 2) 
\frac{\partial u}{\partial x^{\mu}} \frac{\partial u}{\partial z^{\nu}} 
\Biggr\} . \qquad
\end{eqnarray}

\section{Discussion}

We have shown that infrared divergences preclude de Sitter invariant
solutions for the propagators of either a minimally coupled scalar 
with mass $M_S^2 \leq 0$, or for a transverse vector with mass
$M_V^2 \leq -2(D\!-\!1) H^2$. (If one includes the longitudinal
part of the vector propagator then infrared divergences occur for 
$M_V^2 \leq 0$.) However, in most cases these infrared divergences are of
the power law type which is automatically subtracted by any regularization 
which is based upon analytic continuation. (We stress that these 
considerations apply as well to the standard technique of continuation from 
Euclidean de Sitter space.) Only the special values of $M_S^2$ and $M_V^2$ 
given in equation (\ref{probs}) result in logarithmic infrared divergences 
which show up in analytic regularization techniques.  Thus one might reach 
the incorrect conclusion that de Sitter invariant propagators exist for all 
scalar and vector masses, except for a few ``singular'' cases. 

That conclusion is wrong because infrared divergences should not be 
renormalized the way one treats an ultraviolet divergence. The appearnace
of an infrared divergence signals that an unphysical assumption has been 
made, and the right response is to identify the problematic assumption 
and modify it. In our case the unphysical assumption is that the universe 
can have been prepared in coherent Bunch-Davies vacuum for arbitrarily 
long wavelength modes. There is no causal process by which this can be
accomplished in the de Sitter geometry. When one assumes either that the
initially super-horizon modes are in some less singular state \cite{AV},
or else that the spatial manifold is compact \cite{TW3}, the resulting
propagators become infrared finite, but not de Sitter invariant.

In each case the true propagator can be written as the naive, de Sitter 
invariant result (defined by dimensional regularization) plus a de Sitter 
breaking infrared correction which is real and obeys the homogeneous
propagator equation. For the scalar propagator we have,
\begin{equation}
i\Delta(x;z) = i\Delta^{\rm dS}(x;z) + \Delta^{\rm IR}(x;z) \; ,
\end{equation}
where $i\Delta^{\rm dS}(x;z)$ is expression (\ref{expansion}) and
$\Delta^{\rm IR}(x;z)$ is given by relations (\ref{series}-\ref{cdef}).
The analogous expression for the transverse vector is,
\begin{equation}
i\Bigl[\mbox{}_{\mu} \Delta^{\rm T}_{\nu}\Bigr](x;z) =
i\Bigl[\mbox{}_{\mu} \Delta^{\rm dS}_{\nu}\Bigr](x;z) +
\Bigl[\mbox{}_{\mu} \Delta^{\rm IR}_{\nu}\Bigr](x;z) \; ,
\end{equation}
where the de Sitter invariant part is defined by expressions (\ref{dSprop})
and (\ref{ansatz}), and the de Sitter breaking terms are given by
equations (\ref{vecans}) and (\ref{Fmath}-\ref{F4exp}). The full vector
propagator, including the logitudinal part, is given by equations
(\ref{fullprop}) and (\ref{DeltaL}).

It might be wondered what physical sense it makes to consider the
propagators of particles with tachyonic masses. First, there is the
mathematical point that they don't possess de Sitter invariant
propagators, despite what one might conclude by erroneously defining
these propagators with some analytic continuation technique. Second, 
there is the important issue of following the time dependent vacuum 
decay which must occur when symmetry breaking takes place during a 
phase of inflation. In this respect the infrared correction terms 
may be quite important, as they sometimes are for the analogous case 
of FRW geometries with constant deceleration \cite{JMPW2}. 

Another application for our propagators is the projection operators 
for higher spin propagators, in which case there are no physical particles
with tachyonic masses to worry about. For example, consider the 
graviton $h_{\mu\nu}$ in exact de Donder gauge,
\begin{equation}
\Bigl(\delta^{\rho}_{\mu} D^{\sigma} - \frac12 D_{\mu} g^{\rho\sigma}\Bigr)
h_{\rho\sigma} = 0 \; . \label{deDonder}
\end{equation}
Just as the source term for the transverse vector propagator 
equation (\ref{VPeqn}) must be consistent with Lorentz gauge
(\ref{transcon}), so too the source term of the graviton propagator
equation must be consistent with (\ref{deDonder}). The resulting
projection operator turns out to involve the full vector propagator,
\begin{eqnarray}
\lefteqn{i\Bigl[\mbox{}_{\mu\nu} \mathcal{P}_{\rho\sigma}\Bigr](x;x') =
g_{\mu (\rho} g_{\sigma) \nu} i \delta^D(x \!-\! x') - \frac1{D \!-\! 2}
\, g_{\mu\nu} g_{\rho\sigma} i\delta^D(x \!-\! x') \nonumber } \\
& & \hspace{2cm} + \frac12 \sqrt{-g(x)} \left\{ \matrix{
D_{\mu} D_{\rho}' \, i [\mbox{}_{\nu} \Delta_{\sigma}](x;x') +
D_{\mu} D_{\sigma}' \, i [\mbox{}_{\nu} \Delta_{\rho}](x;x') \cr
D_{\nu} D_{\rho}' \, i [\mbox{}_{\mu} \Delta_{\sigma}](x;x') +
D_{\nu} D_{\sigma}' \, i [\mbox{}_{\mu} \Delta_{\rho}](x;x') }
\right\} . \qquad \label{proj}
\end{eqnarray}
One can easily check that the de Donder gauge condition,
\begin{equation}
\Bigl[ \delta_{\alpha}^{\rho} D^{\prime \sigma} - \frac12 D_{\alpha}'
g^{\prime \rho\sigma}\Bigr] \, i \Bigl[ \mbox{}_{\mu\nu}
\mathcal{P}_{\rho\sigma}\Bigr](x;x') = 0 \; .
\end{equation}
requires the vector to have mass $M_V^2 = -2(D\!-\!1) H^2$,
\begin{equation}
\sqrt{-g(x)} \, \Bigl[ \square_x + (D\!-\!1) H^2\Bigr] \,
i\Bigl[\mbox{}_{\mu} \Delta_{\nu}\Bigr](x;x') = g_{\mu\nu}
i \delta^D(x \!-\!x') \; . \label{vec1}
\end{equation}
This is not only tachyonic, it actually corresponds to the first of 
the special cases (\ref{Vlogs}) for which the transverse part harbors
a logarithmic infrared divergence, so the problem would show up even
in an analytic regularization technique.

This is all highly relevant to the debate concerning the de Sitter 
invariance of free gravitons \cite{RPW}. It has long been obvious to 
cosmologists that free gravitons cannot be de Sitter invariant because 
they share the same mode functions as massless, minimally coupled scalars
\cite{Grishchuk}. Indeed, the tensor contribution to the primordial
anisotropies of the cosmic ray microwave background derives from
precisely the same infrared singular dependence of these mode 
functions \cite{Starobinsky}. On the other hand, some relativists 
insist that free gravitons must be de Sitter invariant because de
Sitter invariant solutions exist for the propagator when a gauge
fixing term is added to the action \cite{math}. In previous work we have
shown that adding these gauge fixing terms is not valid, owing to the 
linearization instability \cite{MTW,TW7}. Imposing an exact gauge 
condition such as (\ref{deDonder}) should still be all right, but we
have just seen that it leads to an inevitable breaking of de Sitter
invariance through the projection operator. (This same breaking can
be seen as well in noninvariant gauges by adding the appropriate
compensating gauge transformations \cite{Kleppe}.) Antoniadis and
Mottola long ago discovered a similar problem in another gauge \cite{IAEM}. 
Before our current work one might have dismissed these examples as 
``spurious IR divergences for the Feynman propagator in the sense that
the IR divergences are absent for other values of gauge parameters'' 
\cite{Higuchi}. They now appear as just those cases for which infrared 
divergences, that are always present and which always break de Sitter 
invariance, just happen to go from being of the power law type to 
logarithmic, and hence become visible to analytic continuation techniques.
And the correct procedure is not to ignore them or subtract them but 
rather to remove the erroneous assumption of de Sitter invariance.

\newpage
\centerline{\bf Acknowledgements}

This work was partially supported by FQXi Mini Grant \#MGB-08-008,
by FONDECYT grant 3100041, by European Union grant MRTN-CT-2004-512194, 
by Hellenic grant INTERREG IIIA, by European Union Grant 
FP-7-REGPOT-2008-1-CreteHEPCosmo-228644, by NSF grants PHY-0653085 and 
PHY-0855021, and by the Institute for Fundamental Theory at the 
University of Florida.

\end{document}